\documentclass[a4paper,final]{appolb}
\usepackage[utf8]{inputenc}
\usepackage[usenames,dvipsnames]{color}
\usepackage[dvips]{graphicx}
\usepackage{lineno} 
\usepackage[bookmarks]{hyperref}
        \hypersetup{colorlinks,  linktocpage, citecolor=MidnightBlue, filecolor=Blue, linkcolor=MidnightBlue, urlcolor=MidnightBlue}
\newcommand {\pT}       {\ensuremath{p_{\mathrm{t}}}}
\newcommand {\meanpT}   {\ensuremath{\langle p_{\mathrm{t}} \kern-0.1em\rangle}} 
\newcommand {\sqrtS}    {\ensuremath{\sqrt{s}}}
\newcommand {\sqrtSnn}  {\ensuremath{\sqrt{s_{NN}}}}

\newcommand {\dEdx}     {\ensuremath{\mathrm{d}E/\mathrm{d}x }}
\newcommand {\pp}       {\mbox{$\mathrm {p\kern-0.05em p}$}}

\newcommand {\PbPb}     {\ensuremath{\mbox{Pb--Pb} }}
\newcommand {\AuAu}     {\ensuremath{\mbox{Au--Au} }}

\renewcommand {\AA}     {\ensuremath{\mbox{A--A}}}

\newcommand {\meanNpart}   {\ensuremath{\langle N_{\mathrm{part}} \kern-0.1em\rangle}}
\newcommand {\meanNbin}   {\ensuremath{\langle N_{\mathrm{bin}} \kern-0.1em\rangle}}
\newcommand {\Raa}       {\ensuremath{R_{\mathrm{AA}}}}
\newcommand {\mass}     {\mbox{\rm MeV$\kern-0.15em /\kern-0.12em c^2$}}
\newcommand {\tev}         {\mbox{${\rm TeV}$}}
\newcommand {\gev}        {\mbox{${\rm GeV}$}}

\newcommand {\mom}     {\mbox{\rm GeV$\kern-0.15em /\kern-0.12em c$}}
\newcommand {\gmom}   {\mbox{\rm GeV$\kern-0.15em /\kern-0.12em c$}}
\newcommand {\mmass} {\mbox{\rm MeV$\kern-0.15em /\kern-0.12em c^2$}}
\newcommand {\gmass}  {\mbox{\rm GeV$\kern-0.15em /\kern-0.12em c^2$}}
\newcommand {\mmom}  {\mbox{\rm MeV$\kern-0.15em /\kern-0.12em c$}}

\newcommand {\dg}         {\mbox{$\kern+0.1em ^\circ$}}
\newcommand {\Nch}     {\ensuremath{N_{\rm ch}}}

\newcommand{\Kzs}       {\mbox{$\mathrm {K^0_S}$}}

\newcommand{\rmLambdaZ}         {\mbox{$\mathrm {\Lambda}$}}
\newcommand{\rmAlambdaZ}        {\mbox{$\mathrm {\overline{\Lambda}}$}}

\newcommand{\rmXi}      {\mbox{$\mathrm {\Xi}$}}
\newcommand{\rmXiM}     {\mbox{$\mathrm {\Xi^{-}}$}}

\newcommand{\rmAxiP}    {\mbox{$\mathrm {\overline{\Xi}^{+}}$}}

\newcommand{\rmOmega}   {\mbox{$\mathrm {\Omega}$}}
\newcommand{\rmOmegaM}  {\mbox{$\mathrm {\Omega^{-}}$}}

\newcommand{\rmAomegaP} {\mbox{$\mathrm {\overline{\Omega}^{+}}$}}
\newcommand{\rmOmegas}  {\mbox{$\mathrm {\Omega^{-}+\overline{\Omega}^{+}}$}}

\newcommand{\Jpsi}      {\mbox{J\kern-0.05em /\kern-0.05em$\psi$}}

\begin{document}
\eqsec  
%
%
%
\title{Strange and multi-strange particle production at the LHC energies with ALICE%
\thanks{Presented at {\it Strangeness In Quark Matter 2011}, 18-24 September 2011, Cracow}%
}
\author{Boris Hippolyte\footnote{hippolyt@in2p3.fr} for the ALICE Collaboration
\address{Institut Pluridisciplinaire Hubert Curien and Universit\'{e} de Strasbourg, France}
}
%
%
\maketitle
\begin{abstract}
Strange quark and particle production is studied at the LHC with unprecedented high beam
energies in both heavy-ion and proton-proton collisions: on the one hand, strangeness is
used for investigating chemical equilibration and bulk properties; on the other hand,
strange particles contribute to probe different kinematical domains, from the one where
collective phenomena are at play up to the region dominated by pQCD-calculable
processes.
We highlight the suitability of the ALICE experiment for this topic, presenting our latest
measurements and comparing them to models.
\end{abstract}
%
%
\section{Introduction}
\label{sec:Intro}
Strange particles have been commonly used as probes of the quark--gluon plasma (QGP)
by many experiments at the AGS, SPS and RHIC~\cite{Blume:2011sb}. Several measurements
of strangeness were expected with eagerness at the Large Hadron Collider (LHC)~\cite{Alessandro:2006yt}.
The ALICE experiment recorded high-statistics data in 2010 for both proton-proton and heavy-ion
collisions~\cite{Prino:2011fu} in order to tackle strange and multi-strange hadron production in
detail.

Benefiting from the lowest material budget in the central rapidity region at the LHC, as well as
from complementary particle identification techniques~\cite{Hippolyte:2009xz}, ALICE can measure
most of strange mesons and baryons, weakly decaying or resonances, over a large transverse
momentum ($\pT$) range at the highest colliding energies.
Instead of discussing extensively the wealth of results obtained less than a year after the
first $\AA$ collisions at the LHC, this contribution focus on chosen highlights confirming the behaviours
seen at lower energies and those for which heavy-ion and $\pp$ measurements are strikingly different.
In the next (second) section, we present strange hadron $\pT$ spectra measured in $\pp$
collisions and draw comparisons with spectra from Monte Carlo generators.
Although the description of some aspects of elementary hadronic collisions constitutes 
a challenge, especially for multi-strange baryons, these measurements set the baseline for
$\PbPb$ collisions.
The third section is dedicated to strange baryon/meson ratio as a function of $\pT$, which is
not only important for testing the validity of coalescence models for $\AA$ collisions, but also
helps define the kinematical domain where pQCD-calculable processes dominate.
The excitation function of the multi-strange baryon enhancement from SPS to LHC energies
is discussed in the fourth section.
In the fifth section, the suppression of strangeness spectra in central $\PbPb$ collisions with
respect to that in $\pp$ is put in context of the open-charm measurements.
%
%
\section{Strange and multi-strange $\pT$ spectra in pp collisions at 7 $\tev$}
\label{sec:pt_pectra_pp}
The design of the ALICE experiment was optimized for particle identification (PID), especially
in the central barrel~\cite{Alessandro:2006yt}.
During the first two years of data taking, the complementarity of the detectors was exploited to
obtain high precision $\pT$ spectra in the soft physics regime for both $\pp$ and $\PbPb$ collisions.
In the case of strange hadrons, the identification methods included $\dEdx$ loss in the Inner
Tracking System and the Time Projection Chamber, Time of Flight measurements, as well as
weak decay topology techniques and invariant mass analyses. Several of these methods were
combined so that the production of strange hadrons including hyperons and resonances
could be measured and compared to those from Monte Carlo generators~\cite{Guerzoni2011sqm,Maire2011sqm,Gangadharan2011sqm}.
%
%
\begin{figure}[!htb]
\centering
	\begin{minipage}[c]{.495\linewidth}   
         \includegraphics[scale=0.33, clip=true, trim=0.cm 0.0cm 0.cm 0cm]{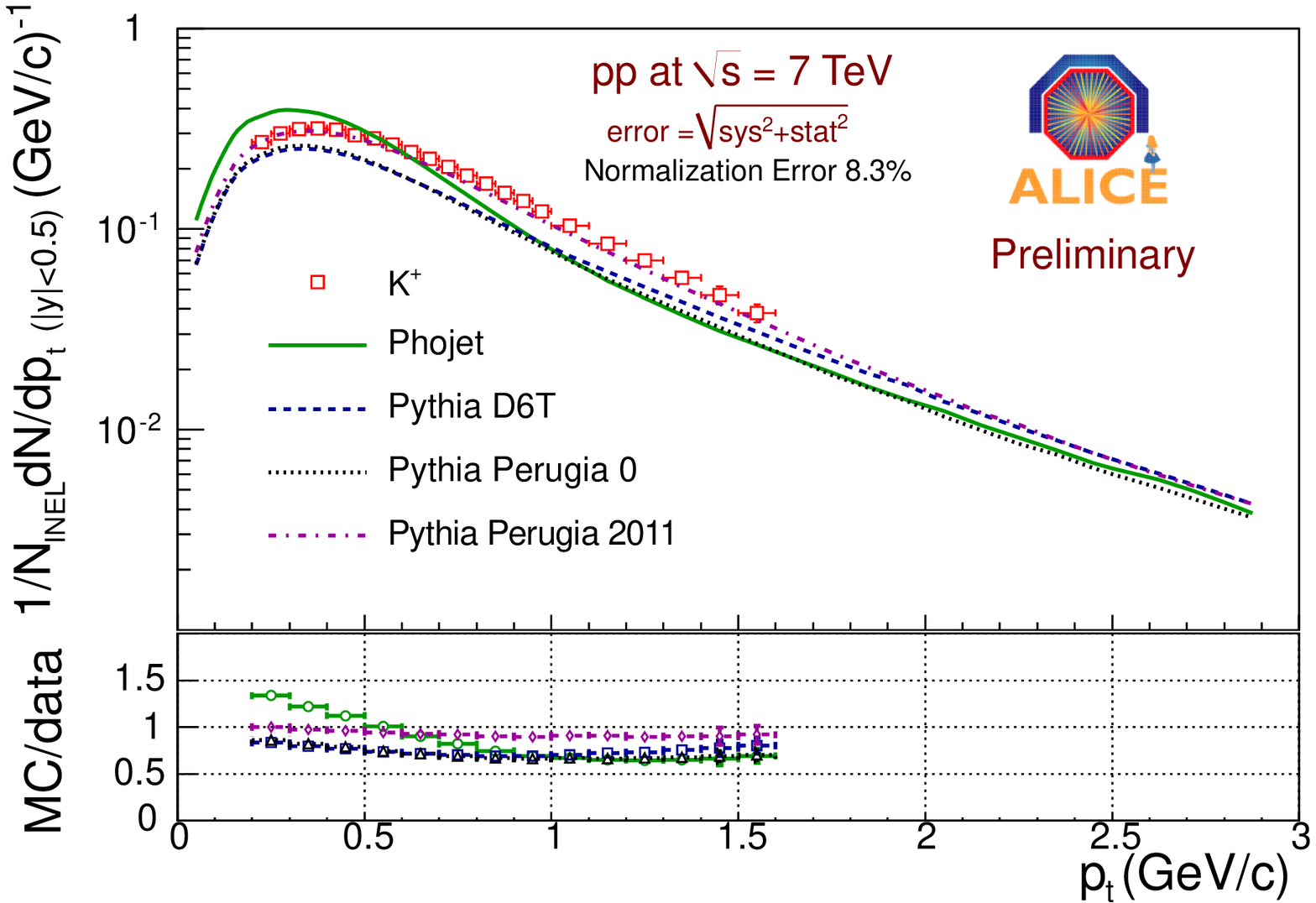}
         \end{minipage} \hfill
         \begin{minipage}[c]{.496\linewidth}
         \includegraphics[scale=0.34, clip=true, trim=0.4cm 0cm 0cm 0cm]{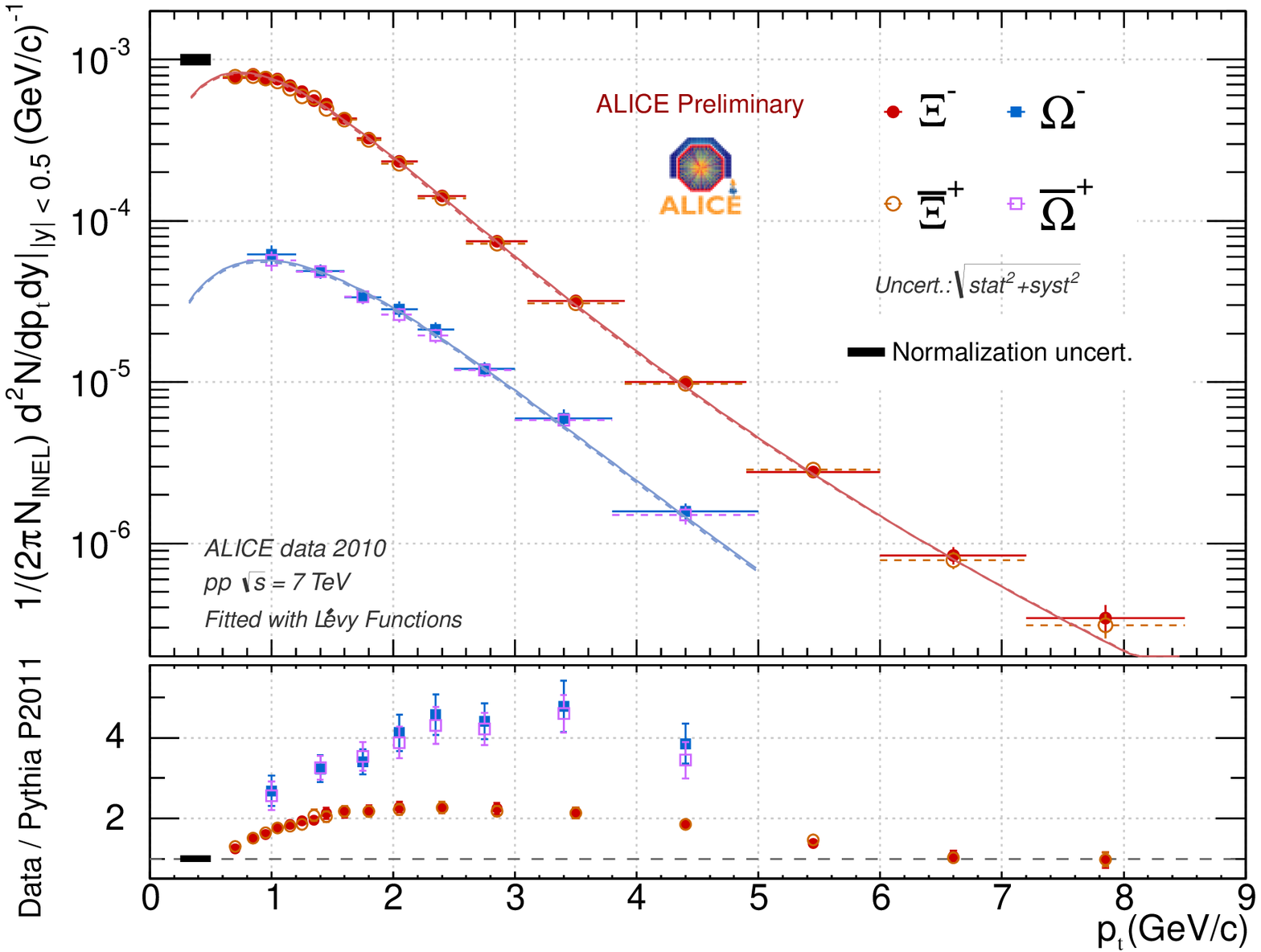}
         \end{minipage}
\caption{Transverse momentum spectra for charged (positive) kaons (left panel)~\cite{Guerzoni2011sqm}
and multistrange baryons ($\rmXiM$, $\rmAxiP$, $\rmOmegaM$ and $\rmAomegaP$, right panel)~\cite{Maire2011sqm}
measured in the central rapidity region ($|y|<0.5$) for pp inelastic events at $\sqrtS=7~\tev$.
Comparisons with PYTHIA 6.4 spectra are also shown.}
\label{fig:pt_spectra_pp_mc}
\end{figure}
%

Figure~\ref{fig:pt_spectra_pp_mc} shows the $\pT$ spectra for charged kaons (left panel) and
multi-strange baryons (right panel: $\rmXiM$, $\rmAxiP$, $\rmOmegaM$ and $\rmAomegaP$)
normalized to inelastic pp events at $\sqrtS=7~\tev$.
Specific details of the analyses are reported in~\cite{Guerzoni2011sqm,Maire2011sqm} where
it is indicated that the L\'{e}vy-Tsallis functional form describes the shape of the $\pT$ distributions
fairly well, as illustrated for hyperons in the right panel.
Consequently, fits with this function are used to extrapolate the measurements down to
$\pT=0~\gmom$ and to extract the global integrated yields for each species, as was done
at $\sqrtS=0.9~\tev$~\cite{Aamodt:2011zj,Aamodt:2011zza}.

Adjusting Monte Carlo generators in $\pp$ (keeping in mind that $\pp$ interactions are
useful as a reference to heavy-ion collisions) is a difficult task. The dedicated efforts of the
authors must be acknowledged, especially for strangeness production~\cite{Skands:2010ak}.
In the years preceding the LHC start-up, these models included phenomenological approaches
such as multi-parton interactions or color reconnections in the case of PYTHIA 6.4~\cite{Sjostrand:2006za}.
These led to a general improvement for mesons, and PYTHIA Perugia 2011 gives a good
description of kaons production in $\pp$ at $\sqrtS=7~\tev$ in particular~\cite{Guerzoni2011sqm}.
It matches the spectra better than other PYTHIA tunes although it is not perfect for multistrange
baryons: spectra are underestimated up to a factor of 2 and 5 for $\rmXi$ and $\rmOmega$
respectively in the intermediate $\pT$ region.
At high $\pT$, the discrepancy seems to decrease for $\rmXiM$ and $\rmAxiP$ when entering
the fragmentation regime. More statistics is needed before trying to draw a similar
conclusion for $\rmOmegaM$ and $\rmAomegaP$. A further constraint is added when the spectra
of baryonic resonances are compared to models~\cite{Gangadharan2011sqm}.
%
%
\section{Strange baryon over meson ratio as a function of $\pT$.}
\label{sec:baryon_meson_ratio}
Many of the aforementioned discrepancies between PYTHIA and the $\pT$ spectra measured
for $\pp$ collisions at the LHC are pertinent to baryons.
From the point of view of Lund's fragmentation the baryons are energetically more difficult to
produce than mesons since baryon creation involves the formation of di-quark pairs instead of quark pairs.
This is partly the reason why the observation at RHIC of a baryon/meson ratio close to unity
in the intermediate $\pT$ region (around $\pT=2.5~\gmom$) for nucleus-nucleus collisions
was unexpected~\cite{Lamont:2007ce}.
This measurement suggested the presence of other hadronization mechanisms, including
possible interplays between soft and hard processes.
Quark coalescence, which could explain qualitatively such a behaviour in $\AA$, was invoked;
in pp collisions, this hadronization scenario is unlikely due to the low phase space
density in the final state.
At the LHC, the follow-up was two-fold:
i) check if the unprecedented centre-of-mass energy would be insufficient for producing
the same effect in $\pp$ collisions with the increased phase-space density;
ii) extract the centrality dependence of the ratio in heavy-ion collisions and then compare the
LHC values to the RHIC ones.
Preliminary measurements were presented in~\cite{Belikov2011qm}, here we only report
the conclusions of the study.
\begin{figure}[!htb]
\centering
	\begin{minipage}[c]{.495\linewidth}   
         \includegraphics[scale=0.33, clip=true, trim=0cm 0cm 0cm 0cm]{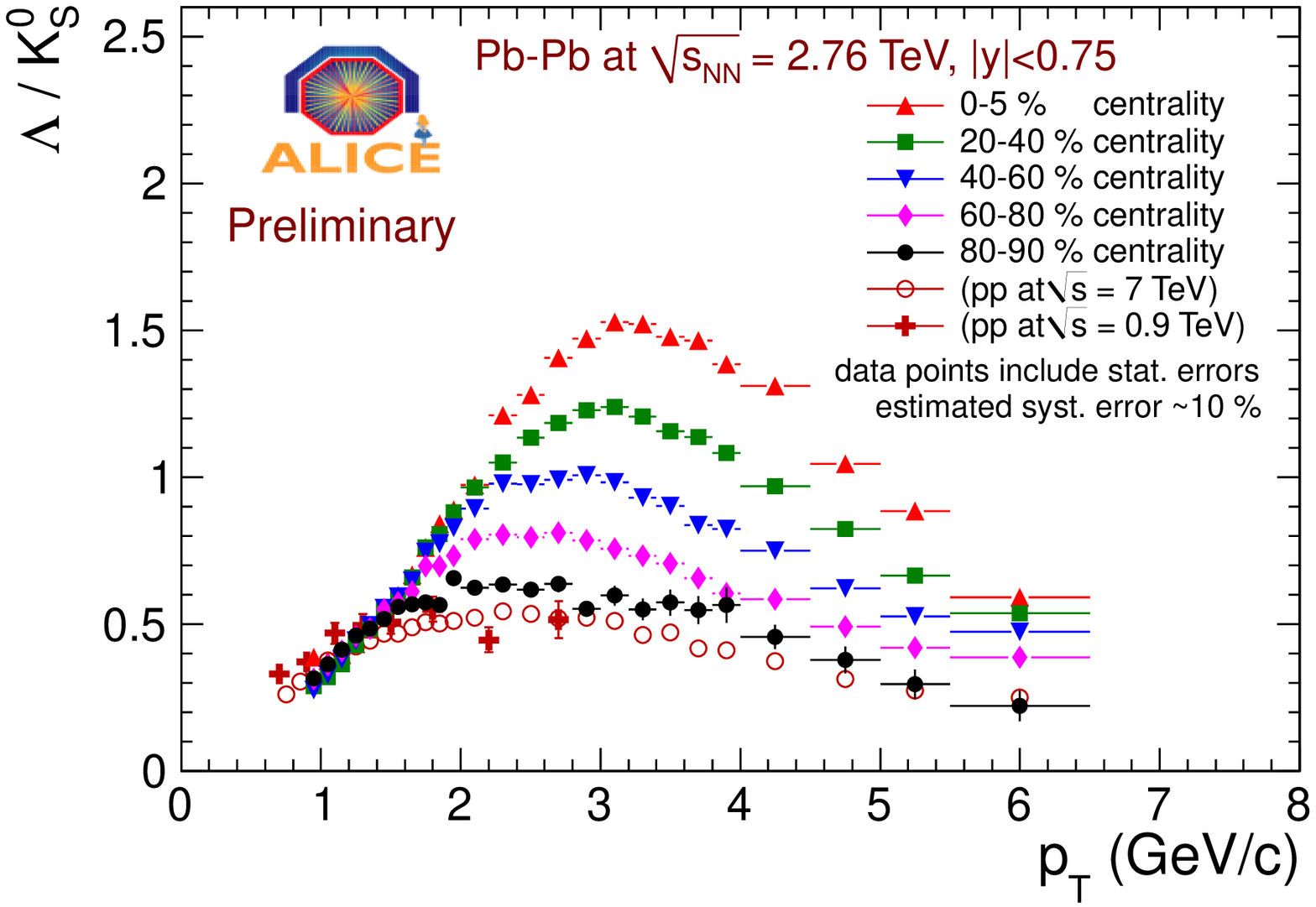}
         \end{minipage} \hfill
         \begin{minipage}[c]{.495\linewidth}
         \includegraphics[scale=0.32, clip=true, trim=0cm 0cm 0cm 0cm]{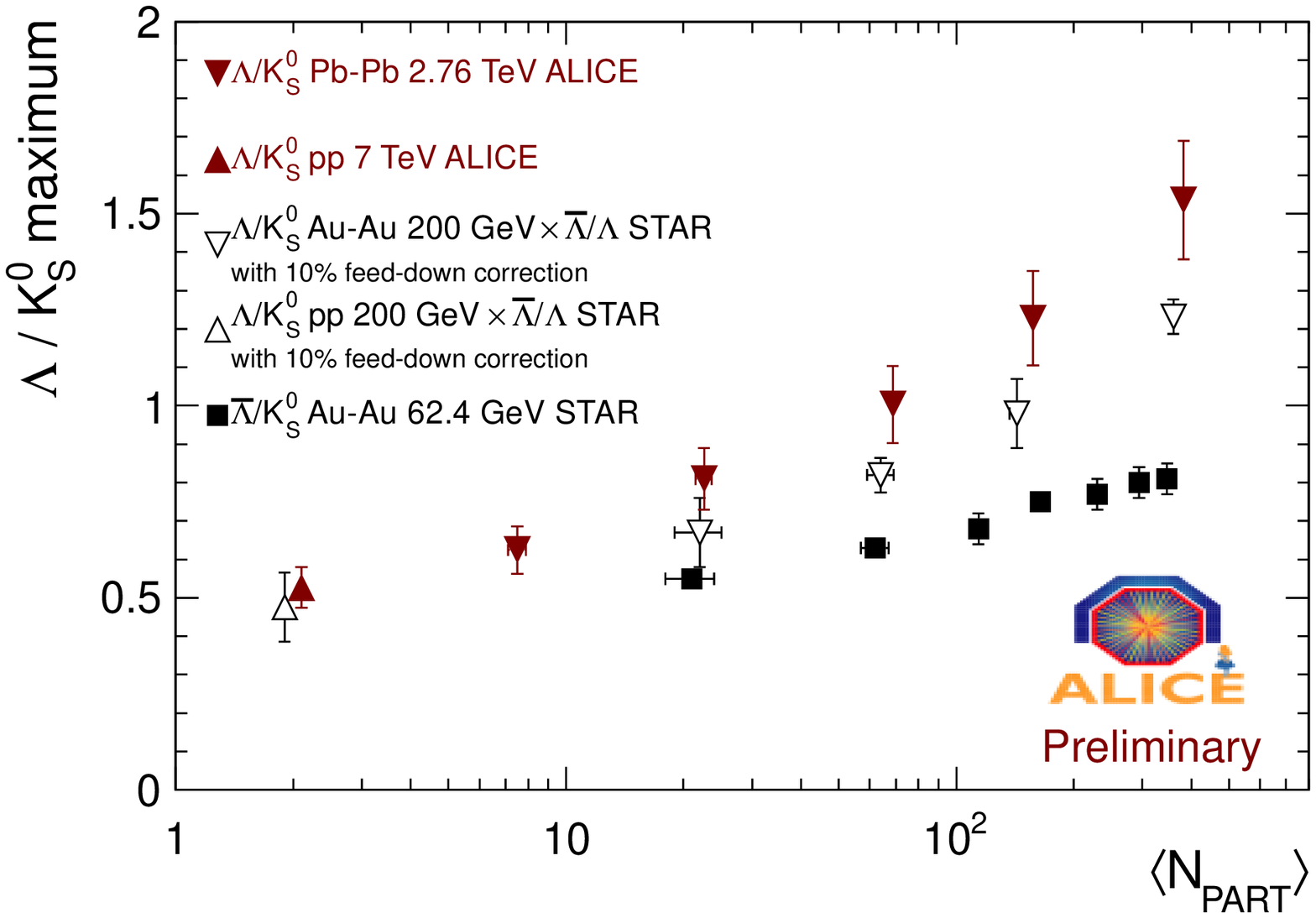}
         \end{minipage}
\caption{Centrality dependence of the baryon over meson ratio as a function of $\pT$
in $\PbPb$ collisions at $\sqrtSnn=2.76~\tev$, illustrated with $\rmLambdaZ/\Kzs$ and
compared with minimum bias pp collisions at $\sqrtS=0.9$ and $7~\tev$ (left panel).
Maximum value of the $\rmLambdaZ/\Kzs$ ratio as a function of the mean number
of participating nucleons ($\meanNpart$) for different colliding systems and energies
(right panel).
Figures are taken from~\cite{Belikov2011qm}. Feed-down corrections are applied.
Because the net-baryon free region is not yet achieved in central rapidities of RHIC collisions,
the $\rmAlambdaZ/\Kzs$ is used for comparison (the $\rmLambdaZ/\Kzs$ is scaled with
$\rmAlambdaZ/\rmLambdaZ$ when appropriate).}
\label{fig:lambda_over_kzs}
\end{figure}
%

The comparison of baryon and meson productions is illustrated by the $\rmLambdaZ/\Kzs$
ratio in Fig.~\ref{fig:lambda_over_kzs}. The measurements are based on a topological identification
method which offers the advantage of covering the relevant $\pT$ range ($1<\pT<6~\gmom$).
Feed-down corrections are performed so that the ALICE preliminary values and STAR results
are directly comparable.
For the pp colliding system, no difference can be seen from $\sqrtS=0.2~\tev$ (by STAR), to $0.9$
and $7~\tev$ (by ALICE): the maximum remains below unity independently of centre-of-mass energy.
While the most peripheral $\PbPb$ collisions exhibit a ratio similar to $\pp$, the value increases with
centrality up to $1.5$ at $\pT \simeq 3~\gmom$ for the $0-5\%$ top central collisions.
The behaviour appears to be specific to $\AA$ with both a centrality and an energy
dependence, as shown in the right panel of Fig.~\ref{fig:lambda_over_kzs}.
%
\section{Excitation function of multi-strange baryon enhancement from SPS to LHC energies.}
\label{sec:msPbPb_enhancement}
An enhanced production of strange hadrons was suggested 30 years ago as an indication
that the state of matter produced in ultra-relativistic nucleus-nucleus collisions is different 
from a hadron gas created at the same energy in nucleon-nucleon collisions~\cite{Rafelski:1982pu}.
Indeed, the large yields found in heavy-ion collisions at the SPS~\cite{Andersen:1999ym,Antinori:2002qr}
and RHIC~\cite{Abelev:2007xp}, especially for multi-strange baryons, could hardly be
understood as coming from a hadronic phase: instead they would require a fast equilibration
and a large correlation volume~\cite{Koch:1986ud}.
The observed effect matches one of the earliest predictions made for the QGP formation,
as it was reiterated at this conference (dedicated discussions can be found in~\cite{Muller2011sqm}
and references therein).
\begin{figure}[!htb]
\centering
	\includegraphics[scale=0.45, clip==true, trim=0cm 0cm 0cm 0cm]{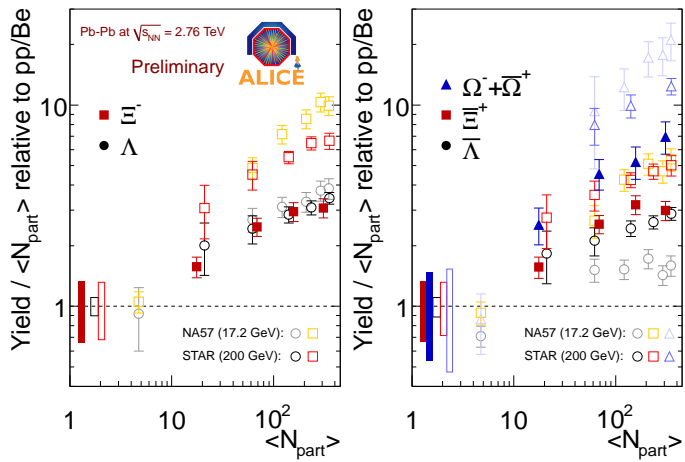}
\caption{Enhancement for hyperon $\PbPb$ yields measured at central rapidity ($|y|<0.5$) with ALICE,
normalized to $\meanNpart$ and using minimum bias $\pp$ values as reference. The left (right) panel
shows baryons which can (not) include valence quarks from the incoming nucleons.
The ALICE results are compared to SPS and RHIC data~\cite{Nicassio2011sqm}. The quadratic sum of
statistical and systematics uncertainties are represented with vertical errors on data points. Uncertainties
on the pp references (or pBe for the SPS data) are noted with rectangles on the left hand of the dotted
line located at unity.}
\label{fig:enhancement_PbPb}
\end{figure}
%

%
Figure~\ref{fig:enhancement_PbPb} summarizes the enhancement factors obtained for different
hyperon species as a function of the mean number of nucleons participating in the collision
for WA97/NA57 at the SPS~\cite{Andersen:1999ym,Antinori:2002qr}, STAR at RHIC~\cite{Abelev:2007xp}
and ALICE at the LHC~\cite{Nicassio2011sqm}.
The corresponding colliding energies are $\sqrtSnn=17.2~\gev$, $200~\gev$ and $2.76~\tev$
respectively. The measurements were performed using topological identification techniques,
common to all experiments and essentially the same as was described in section~\ref{sec:pt_pectra_pp}
for $\pp$ events. 
It must be noted that NA57 results are normalized to pBe yields. The associated enhancements
are calculated for pPb and different centrality classes for Pb-Pb, translated to a mean number of
participating nucleons. No enhancement is seen for pPb contrarily to Pb-Pb.
The STAR and ALICE yields for $\AuAu$ and $\PbPb$ respectively were also extracted as a
function of centrality and translated to a mean number of participants, then normalized
to $\pp$ collisions (in the case of ALICE, the $\pp$ yields are interpolated between $\sqrtS=0.9~\tev$
and $7~\tev$).
Details on the signal extraction and efficiency corrections at $\sqrtSnn=2.76~\tev$ with the ALICE
experiment can be found in~\cite{Nicassio2011sqm}.
This enhancement increases with centrality, and gets systematically larger with
the strangeness content of the hyperon: the maximum value is $\sim$20 for
$\rmOmegas$ in the most central collisions recorded by NA57.
Focusing on the excitation function from $\sqrtSnn=17.2~\gev$ to $2.76~\tev$,
the relative enhancements seem to decrease with increasing energy, as has
been observed at the SPS and then between the SPS and RHIC, even with the
uncertainties on the normalization (depicted as rectangles at the bottom left of
Fig.~\ref{fig:enhancement_PbPb}).
It must be stressed that the absolute production of hyperons in heavy-ion
collisions increases with energy from the SPS to the LHC.
However, the increase of the yields for the systems (pBe or $\pp$) used as
reference, appears to be slightly faster, suggesting that the normalized
heavy-ion yields could saturate.
%
\section{Strangeness suppression at high $\pT$.}
\label{sec:raa_strange}
A spectacular difference between nucleus-nucleus and pp collisions was observed for the
first time at RHIC: the hadron production at large $\pT$ for central $\AuAu$ collisions at
$\sqrtSnn=200~\gev$ is a factor $\sim$4 suppressed when compared to expectations from
a geometrical superposition of nucleon-nucleon collisions~\cite{Adcox:2001jp,Adler:2002xw}.
It was promptly interpreted as resulting from the energy loss suffered by partons produced
in hard parton-scattering when traversing the hot and dense matter created in ultra-relativistic
heavy-ion collisions.
The RHIC observations were confirmed by the first $\PbPb$ collisions at the LHC with
single-particle nuclear modification factor $\Raa$ as a function of $\pT$~\cite{Aamodt:2010jd}.

The aim now is to characterize further the medium properties using the high statistics available
at the LHC and to check whether there is any obvious dependence on the kind of parton probing
the QCD medium.
The suppression is studied as a function of the event centrality in order to extract more information
about the density of the medium; it is observed that i) the amount of suppression increases with
increasing centrality; ii) for all centralities,  the $\Raa$ exhibits a minimum at $\pT =6-7~\gmom$
then iii) it increases slowly up to $\sim$30~$\gmom$~\cite{Otwinowski2011sqm}.
A larger energy loss is also found when comparing the ALICE results to the ones at RHIC: for the
most central ($0-5\%$) collisions, the $\Raa$ measured at the LHC is smaller than that at RHIC,
confirming the expectation that the density of the medium created in the collision increases with
the centre-of-mass energy.
%
\begin{figure}[!htb]
\centering
	\begin{minipage}[c]{.495\linewidth}   
         \includegraphics[scale=0.32, clip=true, trim=0cm 0cm 0cm 0cm]{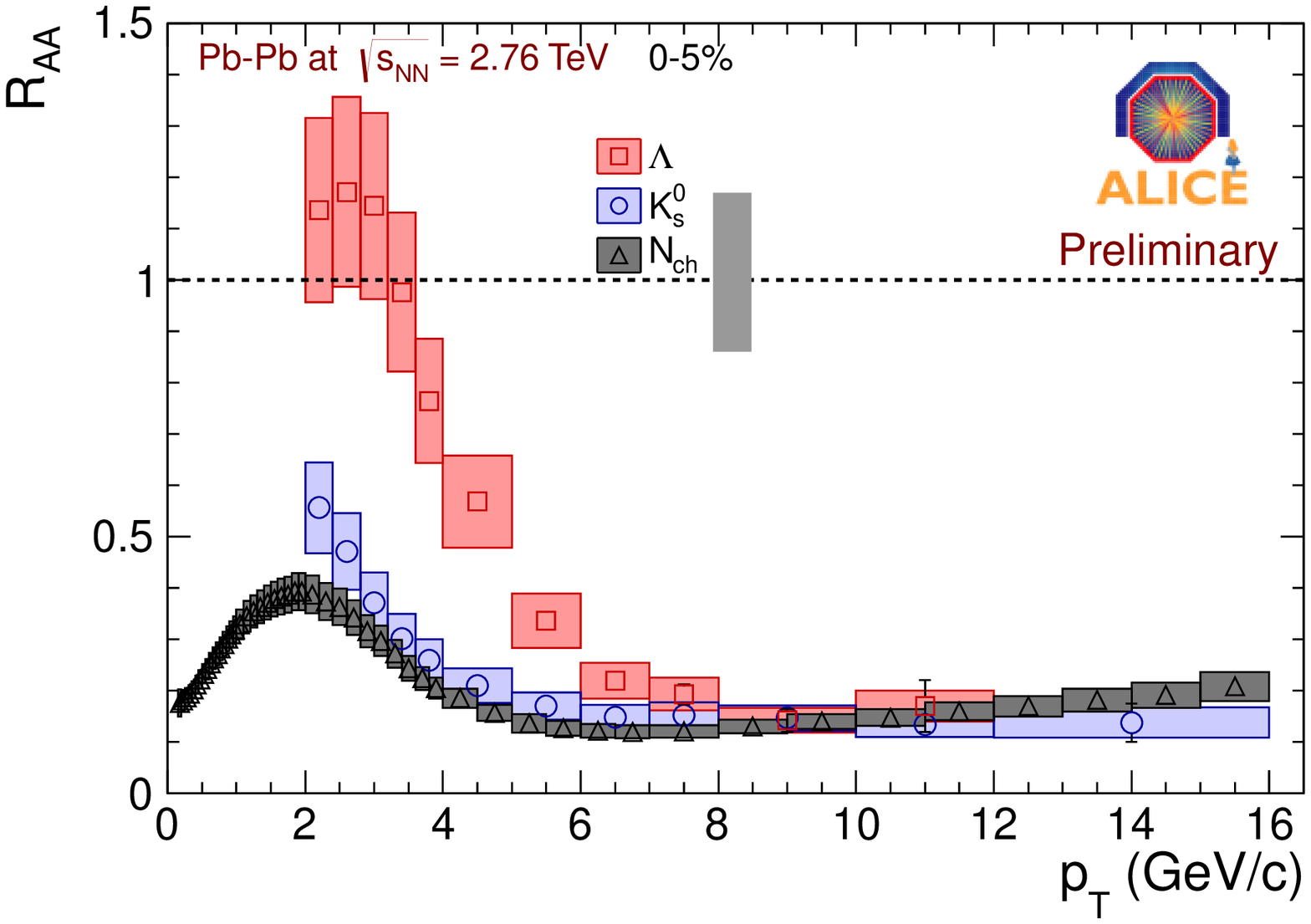}
         \end{minipage} 
         \begin{minipage}[c]{.495\linewidth}
         \includegraphics[scale=0.32, clip=true, trim=0cm 0cm 0cm 0cm]{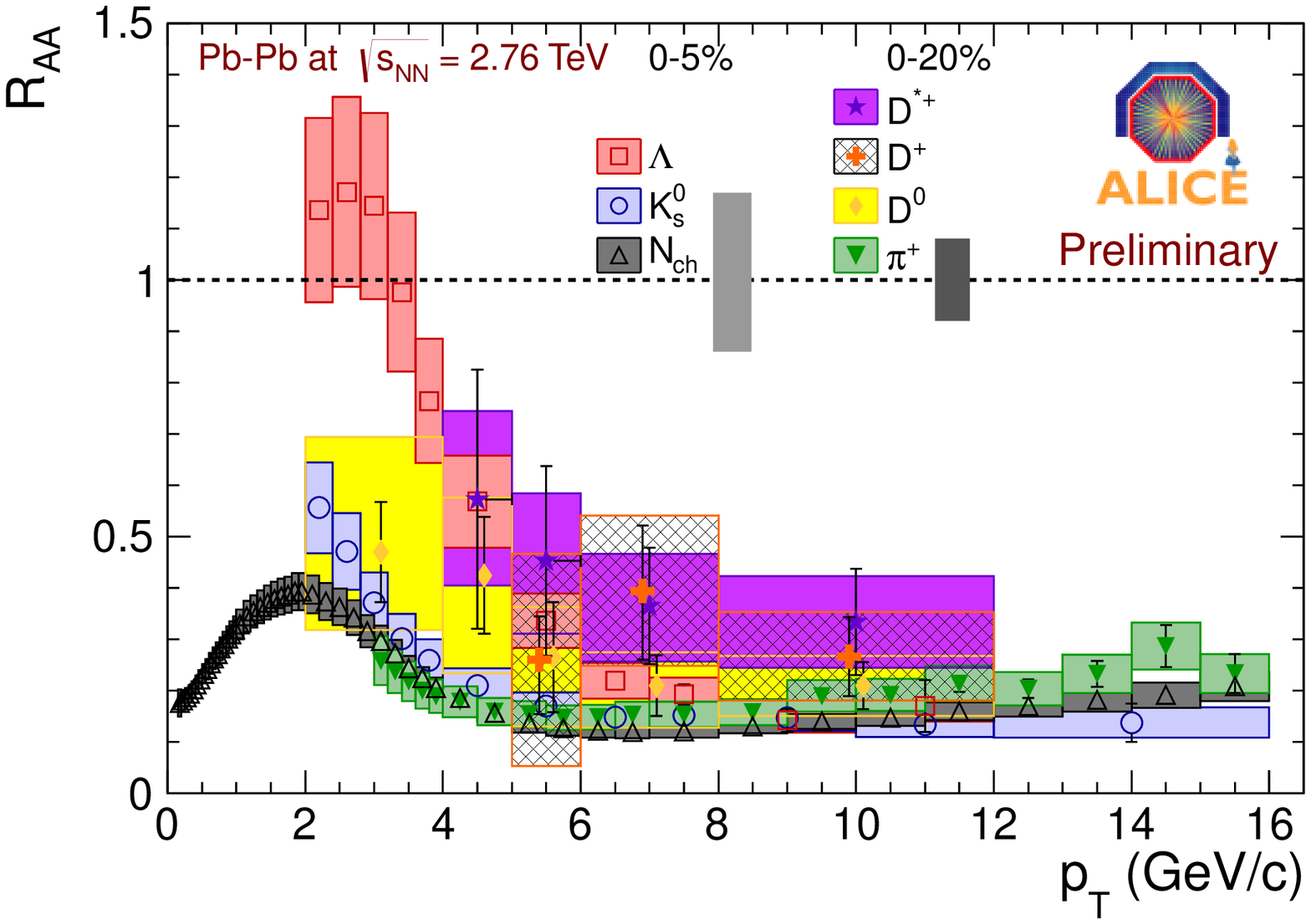}
         \end{minipage}
\caption{The nuclear modification factors $\Raa$ as a function of $\pT$ for $\Kzs$ (circles) and $\rmLambdaZ$
(squares) in central (0--5\%) $\PbPb$ collisions (left panel) together with $\Nch$, the unidentified charged hadrons
(triangles) at $\sqrtSnn=2.76~\tev$. Vertical errors are statistical whereas the boxes corresponds to systematics.
The uncertainty due to the calculation of the mean number of binary collisions ($\meanNbin$) is given by the gray
boxes on the dotted line located at unity~\cite{Schuchmann2011sqm}.
The nuclear modification factors for charmed mesons are superimposed on the right panel for comparison together
with charged pions (using a larger 0--20\% centrality bin was necessary because of statistics).}
\label{fig:raa_central_strange_charm}
\end{figure}

In radiative energy loss models, gluons are expected to loose more energy than quarks
when traversing the hot QCD matter. With the fraction of hadrons from gluon jets increasing
with the colliding energy, it is also possible that the more pronounced suppression seen at
the LHC with respect to RHIC originates from a stronger contribution of gluons.

Figure~\ref{fig:raa_central_strange_charm} also addresses the parton dependence of energy
loss from the flavour point of view~\cite{Schuchmann2011sqm}. On the left panel, it can be
seen that the suppressions of $\Kzs$ and $\rmLambdaZ$ are similar to that of unidentified
charged hadrons for $\pT>6~\gmom$ in the most central (0--5\%) $\PbPb$ collisions.
This is also the case between $\Kzs$ and charged hadrons at lower transverse momenta,
however, the $\Raa$ of $\rmLambdaZ$ is significantly higher and should be related to
alternative hadronization mechanisms as discussed in section~\ref{sec:baryon_meson_ratio}.
A stronger constraint is added to energy loss models when complementing the picture (right panel
of Fig.~\ref{fig:raa_central_strange_charm}) with charmed mesons. Within uncertainties, and
taking into account that a wider centrality had to be used because of statistics, no strong
difference between light and heavy flavour is observed~\cite{Grelli2011sqm}.
%
\section{Summary}
\label{sec:Ccl}
We present the results of several studies illustrating how strangeness production can help
the investigation of the properties of the strongly interacting matter created at the LHC in
heavy-ion collisions at $\sqrtSnn=2.76~\tev$. Measurements are based on the high-statistics
data recorded in 2010 with ALICE.
After setting the $\pp$ baseline with strange hadron $\pT$ spectra, the $\rmLambdaZ/\Kzs$
ratio as a function of $\pT$ is discussed in the context of similar measurements at lower $\sqrtSnn$. 
The $\AA$ ratio seems to be energy-dependent contrarily to the one in pp.
Then the excitation function of hyperons enhancements from SPS to LHC energies is described:
the relative enhancements seem to decrease with increasing energy, confirming the trend observed at
the SPS and then between the SPS and RHIC.
Finally, we report that no strong difference between light and heavy flavour is
observed within uncertainties in central $\PbPb$ collisions for the nuclear modification factor.
%

%
%
\end{document}